\documentclass{PoS}
\usepackage{graphicx}

\makeatletter
\newsavebox{\@parc@ption}
\def\parcaption#1{%
\sbox{\@parc@ption}{\shortstack[l]{#1}}%
\setbox\@tempboxa\hbox{\csname fnum@\@captype\endcsname}%
\@tempdima\columnwidth \advance\@tempdima-\wd\@tempboxa
\@tempdimb.8\@tempdima 
\ifdim\wd\@parc@ption>\@tempdimb \@tempdima\@tempdimb
\else\@tempdima\wd\@parc@ption\fi
\sbox{\@tempboxa}{\parbox[t]{\@tempdima}{#1}}%
\caption{\usebox{\@tempboxa}}}
\makeatother

\title{5-dimensional SU(2) lattice gauge theory with $Z_2$ orbifolding and its phase structure}

\ShortTitle{5-dimensional SU(2) lattice gauge theory with $Z_2$ orbifolding and its phase structure}

\author{Michika Murata\\
 Support Office for Female Researchers Niigata 
University, Ikarashi  2-no-cho 8050, Nishi-ku, Niigata 950-2181, Japan \\
        E-mail: \email{michika@adm.niigata-u.ac.jp}}

\author{\speaker{Hiroto So}%
         \thanks{
         We thank Dr. K.~Ishiyama for the disscusion and basical computations specially. 
         This work was supported in part by Grants-in-Aid for Scientific Research (No.20540274 and  No.21540285) 
 by the Japanese Ministry of Education, Science, Sports and Culture.}\\
           Department of Physics,  Ehime University, Bunkyou-chou 2-5, 
 Matsuyama 790-8577, Japan\\
        E-mail: \email{so@phys.sci.ehime-u.ac.jp}}

\author{Kazunori Takenaga\\
         Kumamoto Health Science
University, Izumi-machi, Kumamoto 861-5598, Japan\\
        E-mail: \email{takenaga@kumamoto-hsu.ac.jp}}

\abstract{In an $SU(2)$ lattice gauge theory  with a $Z_2$ orbifolded extra dimension, 
the new symmetry which is called as a stick symmetry  is useful in understanding 
the bulk transition. We discuss  the  relation with the Fradkin-Shenker's phase diagram 
as well. 
A remnant of the extra dimension is remained as  two $U(1)$ gauge symmetries in two 4-dimensional 
spaces. 
We also consider more general bulk gauge groups beyond  $SU(2)$.}

\FullConference{The XXVIII International Symposium on Lattice Filed Theory\\
		 June 14-19,2010\\
		 Villasimius, Sardinia Italy}

\begin{document}

\section{Motivations}
Besides supersymmetry, there is an attractive scenario called gauge-Higgs 
unification in which the Higgs field is identified with the extra component 
of higher dimensional gauge fields\cite{rev1,rev2}, 
and the gauge symmetry in the 
scenario is caused by the Hosotani mechanism\cite{hoso1,hoso2}.
In the gauge-Higgs  unification scenario by extra-dimensional gauge theories, it is  possible 
to introduce a Higgs field from a  gauge field component naturally and the field  has 
only selfcouplings  originated by the gauge interaction. The potential term is calculable in principle, 
although it is difficult.

Generally, an orbifolding condition is imposed to reduce the dynamical 
freedom of an extra-dimensional system but an origin of 
the symmetry breaking is also generated because the extra-dimensional profile function is 
strongly constrained due to the condition. 
Namely, from orbifolding of an extra-dimensional space, it is possible  to induce 
the origin of the electro-weak symmetry breaking. Since  any perturbative approach assumes 
the existence of a trivial vacuum, although a nonperturbative approach to the unification scenario is valuable,  
the success of its lattice version depends on whether it can describe 
correctly the properties of the vacuum.

In this talk, we investigate  a 5-dimensional $SU(2)$ gauge theory 
with $S^1/Z_2$ orbifolding because $S^1/Z_2$ orbifolding is the simplest way and we follow to  some 
pioneering works for the lattice formulation \cite{I-K1,I-K2,I-K3}. 
The model has a bulk phase transition characterized by the new symmetry 
instead of the center symmetry. The center symmetry is helpless owing to $Z_2$-projection in this case.   
The newly found symmetry is called as a stick one \cite{I-M-S-T}.

A simple method to study the vacuum structure is  to  study the phase diagram which 
 describes the property of the vacuum  based on the investigation of two couplings parameter  space. 
There is a  famous work by Fradkin-Shenker \cite{F-S} about the gauge-Higgs system in 4-dimensional theories. 
Our analysis on the phase diagram is carried forward  comparing  with their work. As the result, we find  that 
two $U(1)$ gauge symmetry in two 4-dimensional spaces are essentially different from any original gauge-Higgs system in 4 dimensions. 

For phenomenological applications, we extend the stick symmetry and find  that 
it is admissible only for bulk gauge groups which has even-dimensional fundamental or defining representations.

\section{Setup and lattice formulation}

We shall start  with a 5-dimensional lattice setup for a large 4-dimensional space($x_{\mu}$) 
and a compact $S^1$($x_5$).  Following to the pioneering works \cite{I-K1,I-K2,I-K3}, 
$S^1  \to S^1/Z_2$ orbifolding is done in the 5th dimension. 
 Although it is natural to take $x_{\mu}=n_{\mu}a$ and 
 $x_5=n_5a_5$, where  the feature of $S^1$ reflects on  the periodic boundary 
 condition along the 5-th coordinate with 
 a size $2L_5$, it is not necessary for $a=a_5$, i.e. an isotropical lattice. 
From the  smoothness of our 4-dimensional space, $a$ is enforced to be extremely small or 
infinitesimal but it is enough for $a_5$ 
to be less than the inverse of our standard model energy scale. 
Our  explicit setup follows as the  gauge group is $SU(2)$, 
a compact 5-dimensional space is $S^1/Z_2$, the size of  our 4-dimensional space is 
  $N_4^4$. 
Accordingly, an $SU(2)$ link variables is described as

\begin{equation}
U_{N,M},~~{ M=1,\cdots 5}, 
\end{equation}
\noindent 
where $N$ is a 5-dimensional lattice coordinate 
$N=\{n_{\nu},n_5\}$ and $M$ indicates the direction.

After $Z_2$ orbifolding($S^1/Z_2$), in a bulk space($n_5=0,\cdots L_5$) 
$SU(2)$ link variables are projected out as follows;
 \begin{eqnarray}
 U_{\{n_\nu  ,n_5 \} ,\mu } & = gU_{\{ n_\nu  ,2L_5  - n_5 \} ,\mu } g^\dag  , \nonumber \\ 
 U_{\{n_\nu  ,n_5 \} ,5}  & = gU^\dag  _{\{ n_\nu  ,2L_5  - n_5  - 1\} ,5} g^\dag  , 
 \end{eqnarray}  
 \noindent
 and for two special points, 
 FP(1):$n_5=0$ and FP(2):$n_5=L_5$,  
 the projections  imply 
 
\begin{equation}\label{eq:U(1)}
 U_{\{n_{\nu}, 0\},\mu} = gU_{\{n_{\nu}, 0\},\mu}g^\dag  , 
 ~~~~U_{\{n_{\nu}, L_5\},\mu} = gU_{\{n_{\nu}, L_5\},\mu}g^\dag ,
\end{equation}
\noindent
where $g = i\sigma _3$.  It is found that these two 4-dimensional spaces, FP(1) and FP(2) and the associated symmetries 
have the important meaning later.  

From a usual action,
\begin{eqnarray}\label{eq:action0}
 S_{S^1 }  = \beta \sum\limits_{P \in M^4  \times S^1 }
  {[1 - \frac{1}{2}{\rm{Tr}}\,\,\,U_P ]} , 
  \end{eqnarray}
\noindent
we obtain our $Z_2$ projected action,   
  \begin{eqnarray}\label{eq:action}
  S_{S^1 /Z_2 }  = 
  \beta  \sum\limits_{P \in bulk} {[1 - \frac{1}{2}{\rm{Tr}}\,\,\,U_P ]} 
   + {\frac{\beta }{2}} \sum\limits_{P \in {\rm FP(1),FP(2)}}
   {[1 - \frac{1}{2}{\rm{Tr}}\,\,\,U_P ]} .
  \end{eqnarray}
\noindent
Here we note the plaquette made of four link variables as $U_P$. 
The second term of (\ref{eq:action}) is written by only $U(1)$ plaquettes from $Z_2$ orbifolding on FP(1) and FP(2) (\ref{eq:U(1)}).  

\section{$Z_2$ orbifolding and Stick symmetry}

For $Z_2$ orbifolded theories, a gauge invariant quantity, is called as $Z_2$-projected 
Polyakov loop, is defined as 

\begin{equation}
L_2 (n_\nu  ) \equiv {\rm{Tr}}\,\,U_{\{ n_\nu  ,0\} ,5} U_{\{ n_\nu  ,1\} ,5} 
\cdots U_{\{ n_\nu  ,L_5  - 1\} ,5} \,\,gU_{_{\{ n_\nu  ,L_5  - 1\} ,5} }^\dag   
\cdots U_{_{\{ n_\nu  ,1\} ,5} }^\dag  U_{_{\{ n_\nu  ,0\} ,5} }^\dag  g^\dag  
\end{equation}
\noindent
where it is noted that every link variable  twice appears in this loop.  
When we apply the usual center symmetry for this loop, this quantity is invariant and 
 this is not suitable for the order parameter of a symmetry breaking. 
Nevertheless, our theory, (\ref{eq:action}) has a phase transition for the loop 
as seen in Fig.1. 
\begin{figure}[htbp]
\begin{center}
\includegraphics[width=7cm]{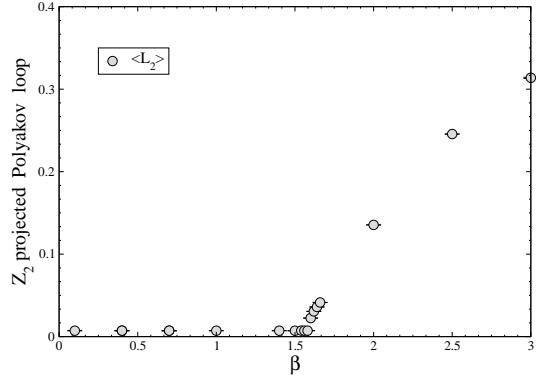}
\caption{VEV of $Z_2$-projected Polyakov loop has been  calculated in the lattice system, $N_4=8,L_5=4$. 
A small hysteresis effect is observed.}
\end{center}
\end{figure}
To understand this situation, in our paper\cite{I-M-S-T}, we proposed a new symmetry(stick symmetry), 

\begin{eqnarray}\label{eq:stick symmetry}
\lefteqn{}U_{\{n_\nu,L_5\},\mu} & \to & i\sigma _2 U_{\{n_\nu,L_5\},\mu}( - i\sigma _2 ) = U^* _{\{n_\nu,L_5\},\mu} , \\
U_{\{n_\nu,n_5\},\mu} & \to & U_{\{n_\nu,n_5\},\mu}~~~{\rm where }~~n_5=0,\cdots,L_5-1 , \nonumber \\
 U_{\{n_\nu,n_5\},5} & \to & U_{\{n_\nu,n_5\},5}~~~{\rm where} ~~n_5=0,\cdots,L_5-2 , \nonumber \\
 U_{\{n_\nu,L_5-1\},5} &\to &  U_{\{n_\nu,L_5-1\},5}( - i\sigma _2 ) . \nonumber
\end{eqnarray}
\noindent
Under the transformation (\ref{eq:stick symmetry}), we can show our action (\ref{eq:action}) and 
the path integral measure are invariant,  
$S_{S^1 /Z_2 }  \to S_{S^1 /Z_2 }$ and  $~~dU_{\{n_\nu,n_5\},M} \to dU_{\{n_\nu,n_5\},M} $. 
Under the stick symmetry, the $Z_2$-projected Polyakov loop is transformed as 
\begin{equation}
L_2 (n_\nu  ) \to - L_2 (n_\nu  ), 
\end{equation}
\noindent
and the VEV can be an order parameter under the symmetry.

\section{Gauge-Higgs system: a comparison with Fradkin-Shenker's phase diagram}
After $Z_2$ projected out, the system is equivalent to a $U(1)^2$ gauge system, $U(1)|_{\rm FP(1)}$ and  $U(1)|_{\rm FP(2)}$ 
on two 4-dimensional spaces(FP(1),FP(2)) and interacting Higgs fields with them.  
That is, in the gauge-Higgs unification, the 5-th component of gauge fields 
corresponds to  Higgs fields and it is possible to compare  with Fradkin-Shenker's 
work\cite{F-S} for gauge-Higgs systems in 4 dimensions. 
Short summary for their work about the phase diagram of the gauge coupling $\beta$ and 
gauge-Higgs coupling $g_h$ is listed in the following. 
For a $U(1)$ gauge plus Higgs system, 
\begin{itemize}
 \item  If a Higgs field has a single-unit charge, the phase diagram has no boundary separated between  
 the confinement and the Higgs phases.  
 \item If a Higgs field has  other charge such as a double unit charge, the phase diagram may have a boundary separated between  
 the confinement and the Higgs phases. 
 \end{itemize}
 \noindent 
 For a non abelian gauge plus Higgs system, 
 \begin{itemize}
 \item If a Higgs field is a fundamental representation, the phase diagram has no boundary separated between  
 the confinement and the Higgs phases.  
 \item If a Higgs field is other representation such as adjoint one, the phase diagram may have a boundary separated between  
 the confinement and the Higgs phases. 
\end{itemize}
\noindent
The case of no boundary separated between  the confinement and the Higgs phases is called as complimentarity.
The important  point of determining  the existence of the phase boundary at $g_h \to \infty$ 
is the vacuum structure of a gauge-Higgs coupling term, 

\begin{equation}\label{eq:higgs}
g_hH^{\dagger}_nU_{n,\mu}H_{n+\hat{\mu}} .
\end{equation}
\noindent
When the Higgs field has a single unit charge or belongs to a fundamental representation, 
 the vacuum configuration of the link variable is unique $U_{n,\mu}\sim 1$.  
 When it has a double unit charge or belongs to an adjoint representation, 
 the vacuum configuration  is $U_{n,\mu}\sim 1$ or $U_{n,\mu}\sim -1$. 
 On the other hand, in the gauge weak coupling limit $\beta \to \infty$  the 
 vacuum configuration is a unique $U_{n,\mu}\sim 1$. Therefore, in the case of 
 a double unit charge or  an adjoint representation, it has a transition point $\beta_{1C}$.  
Here we must note that it is needless to say the gauge vacuum configuration is 
gauge-invariant and it is meaningful only to be unique or doubly degenerate.  
 In the addition  in the gauge weak coupling limit $\beta \to \infty$, 
 the gauge-Higgs interaction is reduses to the 4-dimensional spin model which 
 has a transition point $g_{hC}$. 
 Two typical cases are drawn in Fig. 2 and Fig. 3.

\begin{figure}[htbp]
\begin{minipage}{0.5\hsize}
\begin{center}
\includegraphics[width=7cm]{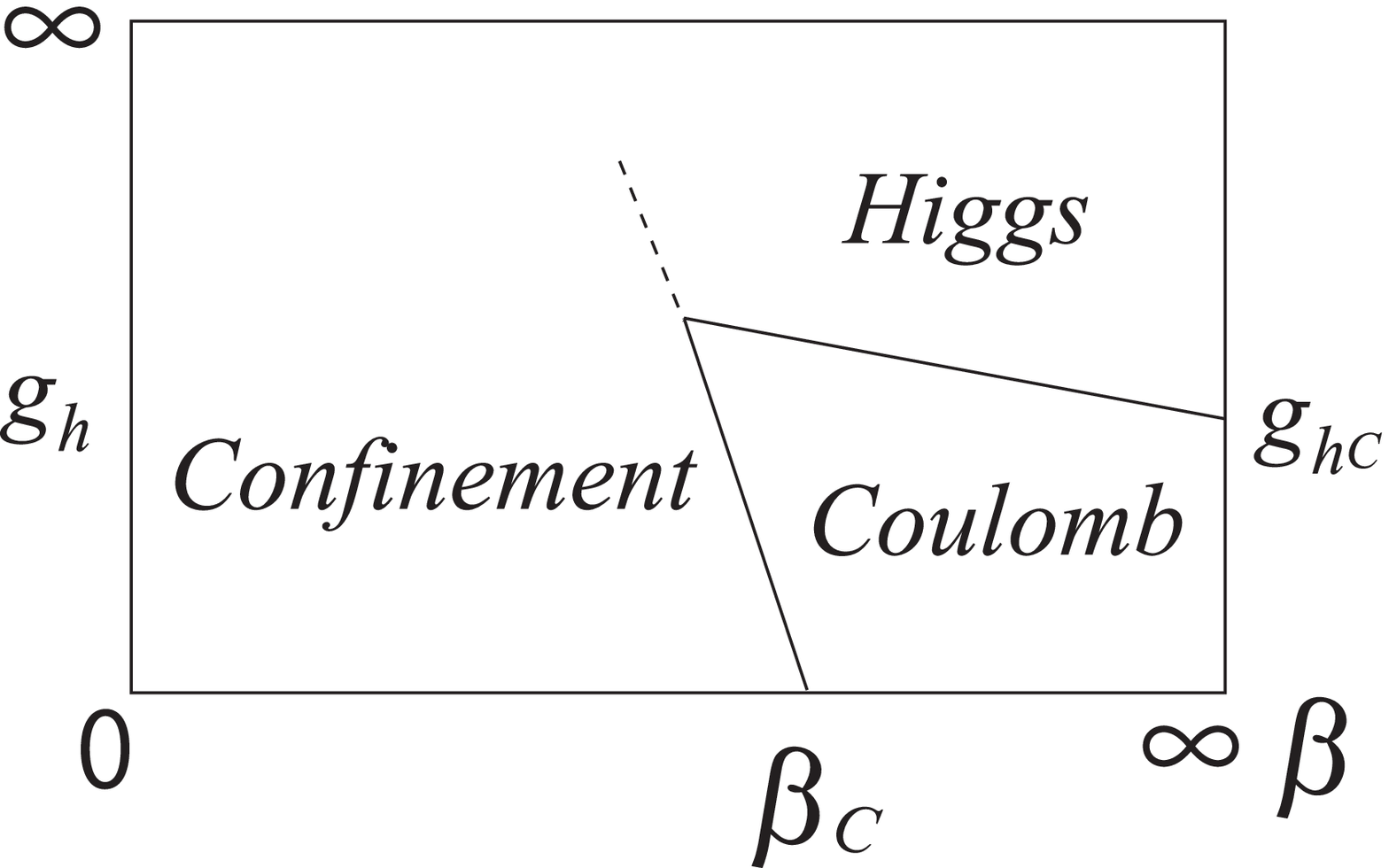}
\parcaption{A  phase diagram of a gauge field and  a fundamental Higgs  field system.}
\end{center}
%
\end{minipage}
\begin{minipage}{0.5\hsize}
\begin{center}
\includegraphics[width=7cm]{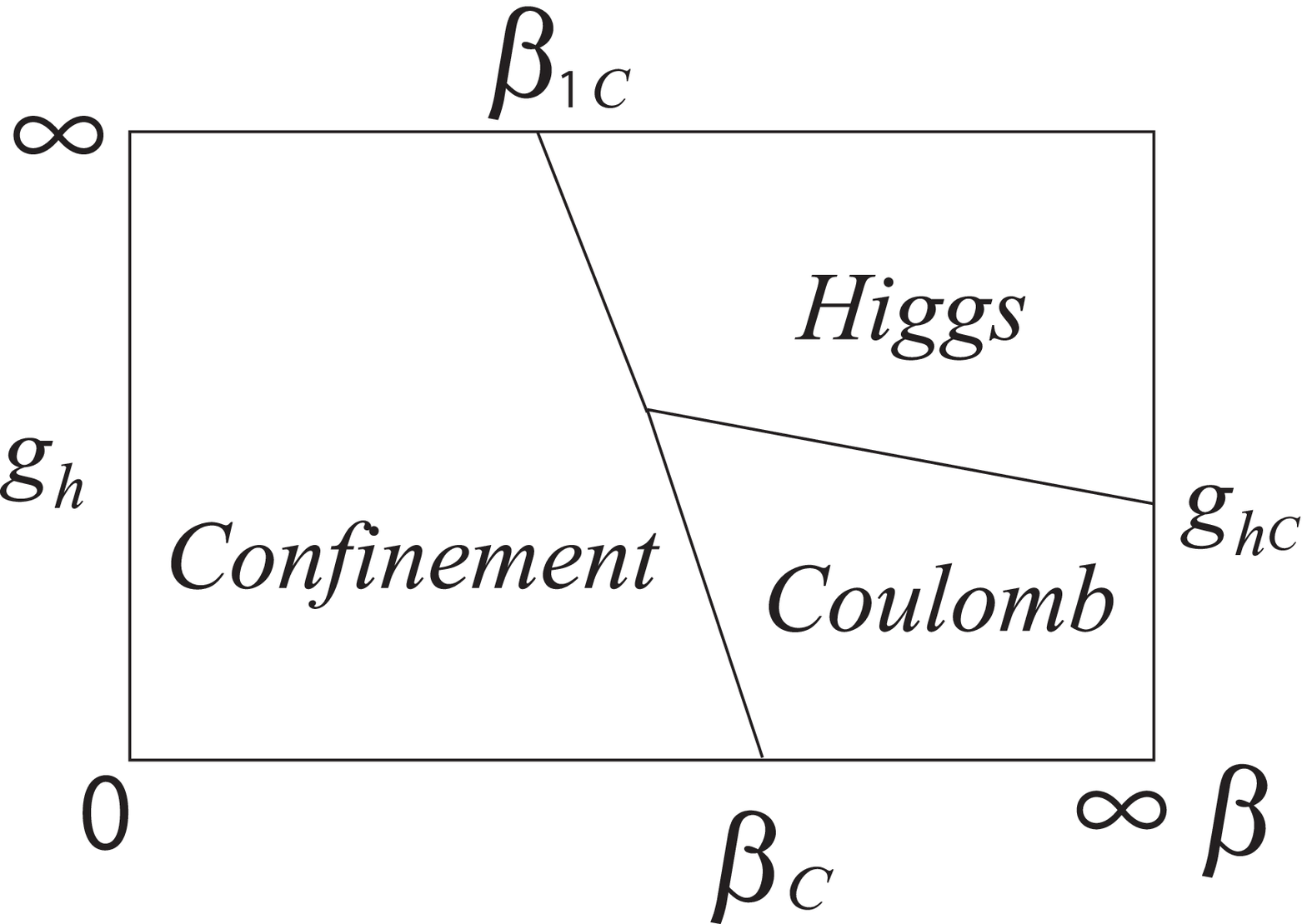}
\parcaption{A  phase diagram of a gauge field and  an adjoint Higgs field system.}
\end{center}
\end{minipage}
\end{figure}
\noindent
When $g_h$  goes to infinity, the theory is equivalent to a 4-dimensional spin model 
coupled to gauge fields with a fixed  vacuum angle.   The vacuum property depends on 
the representation or the magnitude of Higgs fields. 
On the other hand, when $\beta$ goes to infinity, 4-dimensional link variables become  trivial and 
 the resulting theory is equivalent to a usual 4-dimensional spin model. 
 Our system is $U(1) \times U(1)$ gauge fields plus a Higgs field. 
 Apparently, this seems to have a Fig.2-type phase diagram 
 because it has no double unit charge. 
 But, the typical interaction between 4-gauge and 5th-gauge fields is 
 
 \begin{equation}\label{eq:g-h-int}
{\rm Tr~}U_{\{n,L_5-1\},5}U_{\{n,L_5\},\mu}U^{\dagger}_{\{n+\hat{\mu},L_5-1\},5}
U^{\dagger}_{\{n,L_5-1\},\mu} ,
\end{equation}
\noindent
and  the vacuum configuration of the gauge field on the FP(2) 
is lead to $U_{\{n,L_5\},\mu}\sim \pm 1$. Consequently, 
our system has  a similar phase diagram 
to Fig.3.
 
 Although this discussing model is an effective 4-dimensional gauge-Higgs action, 
 our actual model is a 5-dimensional one.   
 This has  so-called a bulk phase transition found  by Creutz 
 \cite{Creutz}.  The remnant of the 5th dimension in the effective theory 
 is remained as  the symmetry $U(1) \times U(1)'$. 
 The difference between two U(1) gauge symmetries is characterized 
 by vaious  extra-dimensional theories such as matter fields. 
   Based on these results, our model is conjected as  Fig.4.

\begin{figure}[htbp]
\begin{center}
\includegraphics[width=7cm]{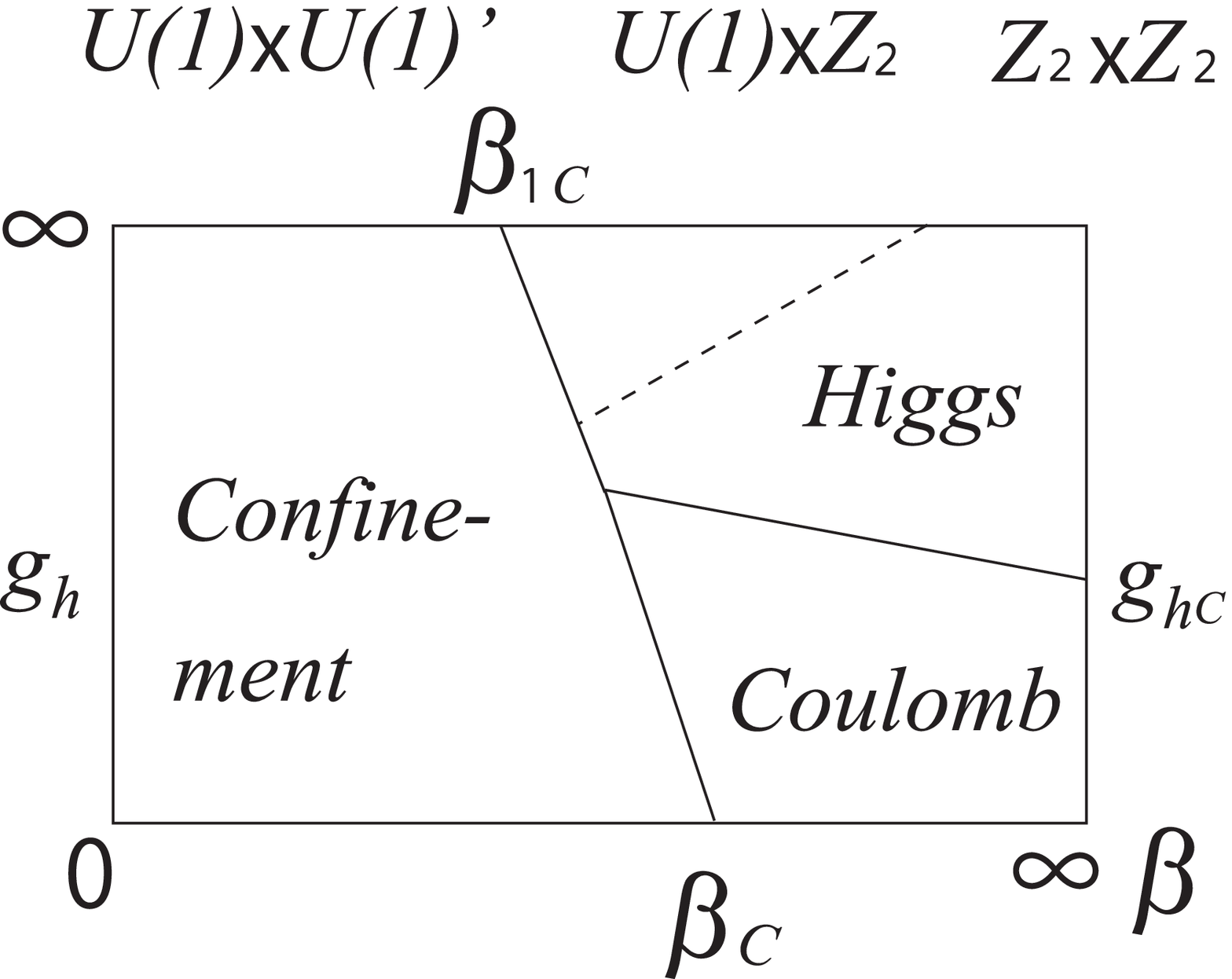}
\caption{~A  phase diagram of a 5-dimensional  gauge system with $Z_2$ orbifolding. 
It is expected  that the dashed line implies a phase boundary 
for the breaking of the remained $U(1)$.  }
\end{center}
\end{figure}
 \noindent

 Instead of the continuum theory, gauge variant variables in a lattice gauge theories 
 have no physical meaning at all such as   profile functions 
of   5-th components in the bulk gauge fields. Since couplings between FP(2) link  variables  
and 5-th components in link variables on the bulk space are gauge invariant,  the Higgs field has a 
 double unit charge coupled to  the FP(2) link variables because of non existence of $U(1)|_{\rm FP(2)}$ monopole. 
 
\section{Generalization of stick symmetry}
We shall consider conditions when there exists the stick symmetry that a bulk gauge symmetry, $G$, breaks into  
a fixed point gauge symmetry, $H$, where $H=\{h\in  G|ghg^\dag=h, g\in G \}$. 
From the property of $Z_2$ orbifolding, 
$g^2$ belongs to the center of $G$, $C(G)$. 
This means the matrix $g$ which describes $Z_2-orbifolding$ can be written as 

\begin{equation}\label{eq:cond1}
g=\Omega^{1/2}V S V^{\dagger}
\end{equation}
\noindent
where $\Omega$ is an element of C(G), $V$ is a unitary matrix, and $\epsilon$ 
is  defined by $S_{ij}=\epsilon_i\delta_{ij}$ with $\epsilon_i=\pm 1$.    
Still, the VEV of $Z_2$-projected Polyakov loop must vanish 
  in the strong coupling limit because of the confinement phase,

\begin{equation}\label{eq:cond2}
<L_2> ~\propto~ {\rm{Tr}}~g=0 .
\end{equation}
\noindent
From (\ref{eq:cond1}) and (\ref{eq:cond2}), the following condition is lead  as 

\begin{equation}\label{eq:cond3}
\sum_i~\epsilon_i =0 , 
\end{equation}
\noindent
and it means the irreducible representation belongs to an even dimensional space. 
 If the bulk gauge theory is made of fundamental or defining representations of 
 $SU(2N),SO(2N),F_4,E_8,USp(2N)$, 
 its stick symmetry can be realized.

\section{Summaries}
We consider a lattice $SU(2)$ gauge theory with $S^1/Z_2$ orbifolding in  5 dimensions.  
From the numerical simulation, it is guessed that there should exist a certain symmetry  in even $S^1/Z_2$ orbifolding.   
The new symmetry is called  as stick one and corresponds to  a kind of a charge conjugation in  a fixed point 4-dimensional space, FP(2). 
Comparing with Fradkin-Shenker's phase diagram, we can conjecture our theory have one or two phase  transition point(s). 
The first transition is the bulk one by found Creutz firstly \cite{Creutz}. The second transition is originated in  
a $U(1)$ gauge theory on a  4-dimensional space(FP(2)).  
If our theory has only a single transition point, it is difficult to take the continuum limit.  
If our theory has two transition points and a $U(1)$ transition continuously 
occurs under the confinement phase in the bulk theory, we expect the theory can take the continuum limit.

Phenomenologically, the bulk $SU(2)$ gauge symmetry  is too small and we extend the symmetry. 
Since  lattice gauge theories necessarily have a confinement phase in the strong coupling region, 
the $Z_2$-projected Polyakov loop  must vanish in the region and it means a stick symmetry unbroken phase. 
 This leads us to a condition; ${\rm Tr~}g=0$.  If the bulk gauge theory is one of 
 $SU(2N),SO(2N),F_4,E_8,USp(2N)$, 
one shall find  its stick symmetry.


\begin{thebibliography}{99}
\bibitem{rev1} C.~Csaki, J.~Hubisz and P.~Meada, 
 {\it TASI Lectures on electroweak symmetry breaking from extra dimensions,} 
   [arXiv:hep-ph/0510275]. 
\bibitem{rev2} H.~C.~Cheng, 
{\it Little Higgs, Non-standard Higgs, No Higgs and all that,} 
[arXiv:0710.3407(hep-ph)]. 
 \bibitem{hoso1} Y.~Hosotani, 
 {\it Dynamical Mass Generation by Compact Extra Dimensions},
 Phys.~Lett.~{\bf B126}~(1983)~309. 
 \bibitem{hoso2} Y.~Hosotani, 
 {\it Dynamics of Nonintegrable Phases and Gauge Symmetry Breaking,}
  ~Ann.~ Phys.{\bf~190}~(1989)~233.
   \bibitem{I-K1} N.~Irges and F.~Knechtli, 
  {\it Non-perturbative definition of five-dimensional gauge theories on the R$^4 \times S^1/Z_2$ orbifold,} 
  Nucl.~Phys.~{\bf B719}(2005)~121~[hep-lat/0411018]. 
  \bibitem{I-K2} N.~Irges and F.~Knechtli, 
  {\it Non-perturbative mass spectrum of an extra-dimensional orbifold,} hep-lat/0604006. 
    \bibitem{I-K3} N.~Irges and F.~Knechtli, 
  {\it Lattice gauge theory approach to spontaneous symmetry breaking from an extra dimension,} 
   Nucl.~Phys.~{\bf B775}(2007)~283~[hep-lat/0609045].
\bibitem{I-M-S-T} K.~Ishiyama, M.~Murata, H.~So and K.~Takenaga, 
  {\it Symmetry and $Z_2$ Orbifolding Approach in Five-dimensional Lattice Gauge Theory,} 
  Prog.~Theor.~Phys.~{\bf 123}~(2010)~257~[arXiv:0911.4555]. 
  \bibitem{F-S} E.~Fradkin and S.~H.~Shenker, 
  {\it Phase diagrams of lattice gauge theories with Higgs fields,} 
  Phys.~Rev.~{\bf D19}~(1979)~3682.  
 \bibitem{Creutz} M.~Creutz, 
{\it Confinement and the Critical Dimensionality of Space-Time,} 
  Phys.~Rev.~Lett.~{\bf 43 }(1979)~553.
   \end{thebibliography}
\end{document}